\documentclass[a4paper,twocolumn,pra,showpacs,aps]{revtex4-1}
\newcommand{\dbar} {\ensuremath{\,\mathchar'26\mkern-12mu d}}

\usepackage{epsfig,amsmath,amsfonts,amssymb}
\usepackage{amsmath,amsfonts,amssymb}
\usepackage{graphicx}
\usepackage{subfigure}
\usepackage{mathbbol}

\begin{document}

\title{Optimal Efficiency of a Noisy Quantum Heat Engine}

\author{Dionisis Stefanatos}
\email{dionisis@post.harvard.edu}
%\altaffiliation[Current address:]{ 3 Omirou St., Sami, Kefalonia 28080, Greece.}
%\email{dionisis@post.harvard.edu}
\affiliation{3 Omirou St., Sami, Kefalonia 28080, Greece}

\date{\today}% It is always \today, today,

\begin{abstract}
In this article we use optimal control to maximize the efficiency of a quantum heat engine executing the Otto cycle in the presence of external noise. We optimize the engine performance for both amplitude and phase noise. In the case of phase damping we additionally show that the ideal performance of a noiseless engine can be retrieved in the adiabatic (long time) limit. The results obtained here are useful in the quest for absolute zero, the design of quantum refrigerators that can cool a physical system to the lowest possible temperature. They can also be applied to the optimal control of a collection of classical harmonic oscillators sharing the same time-dependent frequency and subjected to similar noise mechanisms. Finally, our methodology can be used for the optimization of other interesting thermodynamic processes.
\end{abstract}

\pacs{05.70.Ln, 03.65.Yz, 02.30.Yy, % Control theory
}
%\pacs{05.45.Xt}% PACS, the Physics and Astronomy
\maketitle

\section{INTRODUCTION}

A quantum heat engine executing the Otto cycle is the prototypic quantum system which has been extensively used in the quest for absolute zero, the attempt to cool a physical system towards lower and lower temperatures \cite{Rezek06,Rezek09,Salamon09,Rezek11}. The overall performance of the engine is considerably affected by the necessary time to perform the adiabatic expansion and compression phases of the cycle. Although in general an infinite amount of time is needed to complete an adiabatic operation with perfect fidelity, it has been shown that for the heat engine at hand the same fidelity as with the adiabatic process can be obtained in finite time \cite{Salamon09}. Even shorter times can be achieved using the recently proposed ideas of shortcut to adiabaticity \cite{Chen10,Stefanatos10,Schaff11,shortcuts13} and transitionless quantum driving \cite{Demir08,Berry09,Bason12}, as suggested in \cite{Stefanatos11,Hoffmann_EPL11,Stefanatos13,Deng13,delCampo13}. Note that the perfect, effectively adiabatic evolution is actually an idealization, since in practice there is always present some kind of external noise. In all the above mentioned works, the effect of noise has been considered only indirectly, by requiring to minimize the evolution time, and thus the duration of undesirable exposure to the noise sources. Recently, the effect of noise to the fast adiabatic-like dynamics has attracted some attention and studied for open two-level \cite{Ruschhaupt12,Jing13} and multi-level \cite{Vacanti13} quantum systems, cooling \cite{Choi12}, and for the aforementioned heat engine performing the noisy quantum Otto cycle \cite{TorronKos13}.

In the present article we use optimal control to maximize numerically the performance of the noisy quantum heat engine presented in \cite{TorronKos13}. Note that optimal control has been successfully employed to improve various tasks in the dynamics of open quantum systems. For example, to maximize the efficiency of polarization-coherence transfer between coupled spins in
Nuclear Magnetic Resonance (NMR) \cite{Khaneja03,KhanejaPNAS03,Stefanatos04,Stefanatos05}, to control the relaxation of a qubit \cite{Bonnard12,Lapert13,Mukherjee13} as well as in multi-level quantum systems \cite{Sklarz04,Jirari05,Yuan12,Assemat12}, to maximize the fidelity of quantum gates \cite{Grace07}, and even to manipulate quantum coherence phenomena in light harvesting dynamics \cite{Caruso12}. In parallel with these theoretical works related to the control of noisy quantum systems, there is also considerable experimental progress. As an example we mention the emerging area of quantum optomechanics \cite{Meystre13}, where micro- or nano-mechanical oscillators are coupled to optical cavities \cite{Chan11,Verhagen12}. Full control of the mechanical oscillator quantum state can be achieved using the cavity field, if the coherent coupling rate exceeds the decoherence rate of each subsystem \cite{Verhagen12}. For this kind of noisy quantum systems, the optimal efficiency during adiabatic cooling is an experimental problem of much current interest. Another motivation for the present study is provided by the renewed interest in optimal thermodynamic processes \cite{Salamon09,Hoffmann_EPL11,Schmiedl07,Deffner10,Sivak12,Zulkowski12,Hoffmann13,Deffner13,Deffner14}. Optimal control is the ideal mathematical tool to tackle this kind of problems. Note that the relation between optimal control and thermodynamics is deeper and can be traced back to Carath\'{e}odory. The famous mathematician, with seminal contributions in the calculus of variations which paved the way to optimal control theory \cite{Pesch13}, pioneered the axiomatic formulation of thermodynamics along a purely geometric approach \cite{Caratheodory09}.

In the next section we quickly recall the model of a noisy quantum heat engine proposed in \cite{TorronKos13}. In section \ref{sec:Optimal_solution} we formulate the problem of maximizing the engine efficiency in terms of optimal control and present an appropriate numerical optimization method. This method is used in section \ref{sec:comparison} to obtain the optimal inputs and the corresponding efficiency of the engine. The section also contains a discussion of the results. Section \ref{sec:conclusion} concludes the paper.

%collection parametric oscillators \cite{Hoffmann13}

%the quantum open system as a model of the heat engine\cite{Alicki79}

%Heat engines in finite time governed by master equations\cite{Feldmann96}

%%%%%%%%%%%%%%%%%%%%%%%%%%%%%%%%%%%%%%%%%%%%%%%%%%%%%%%%%%%%%%%%%%%%%%%%%%%%%%%%
\section{QUANTUM OTTO CYCLE WITH EXTERNAL NOISE}

\label{sec:Otto_cycle}

In this article we consider the model of a noisy quantum heat engine proposed in \cite{TorronKos13}. The working medium of the engine is an ensemble of noninteracting particles confined by a harmonic potential with a bounded time varying frequency $\omega_c\leq\omega(t)\leq\omega_h$, which serves as the external control of the system. During the execution of a quantum Otto cycle (where changes in the stiffness of the potential correspond to changes in volume \cite{Rezek06}), the working medium is extracting heat from a cold bath at temperature $T_c$ and delivers it to a hot bath at temperature $T_h$. The cycle is composed by four phases: two isochores, where the medium is in contact with the hot or cold bath while $\omega(t)=\omega_h$ or $\omega(t)=\omega_c$, respectively, and the expansion/compression phases, where the medium is isolated from these baths and $\omega(t)$ is changed from $\omega_h$ to $\omega_c$ or vice versa.

During the operation of the engine, the time evolution of a quantum observable $\hat{A}$ in the Heisenberg picture is given by
\begin{equation}
\label{Observable}
\frac{d\hat{A}}{dt}=\frac{i}{\hbar}[\hat{H},\hat{A}]+\mathcal{L}(\hat{A})+\frac{\partial\hat{A}}{\partial t},
\end{equation}
where
\begin{equation}
\label{Hamiltonian}
\hat{H}=\frac{\hat{p}^2}{2m}+\frac{m\omega^2(t)\hat{q}^2}{2}
\end{equation}
is the Hamiltonian for the working medium with particles of mass $m$ and $\mathcal{L}$ is the Liouville superoperator expressing the effect of noise, which is different for various phases of the cycle \cite{TorronKos13}. At the isochores, this term results to a trivial exponential decay towards thermal equilibrium \cite{TorronKos13}. For this reason we will not pursue further this case but rather concentrate on the more interesting dynamics at the expansion/compression phases. We emphasize that for these phases the engine is isolated from the heat baths of constant temperature but it is subject to fluctuations in the external control which induce noise \cite{TorronKos13}. The corresponding Liouville superoperator is \cite{TorronKos13}
\begin{equation}
\label{Superoperator}
\mathcal{L}(\hat{A})=-\frac{\gamma_p}{\hbar ^2}[\hat{H},[\hat{H},\hat{A}]]-\gamma_a\omega^2[\hat{B},[\hat{B},\hat{A}]],
\end{equation}
where $\hat{B}=m\omega\hat{q}^2/2\hbar$ and $\gamma_p, \gamma_a$ are constants expressing noise strength. As explained in \cite{TorronKos13} and in appendix \ref{app:stochastic}, the first term in (\ref{Superoperator}) corresponds to phase damping \cite{Nielsen00,Feldmann06}, while the second term to random fluctuations in the stiffness of the harmonic potential. The derivation of the master equation (\ref{Observable}) with decoherence given by (\ref{Superoperator}) relies on the assumption that the two noise mechanisms are independent and represent zero mean gaussian white noises, see for example \cite{Rahmani}.

The Hamiltonian (\ref{Hamiltonian}), the Lagrangian
\begin{equation}
\label{Lagrangian}
\hat{L}=\frac{\hat{p}^2}{2m}-\frac{m\omega^2(t)\hat{q}^2}{2}
\end{equation}
and the position-momentum correlation
\begin{equation}
\label{Corellation}
\hat{C}=\frac{\omega(\hat{q}\hat{p}+\hat{p}\hat{q})}{2}
\end{equation}
form a closed set under the time evolution generated by $\hat{H}$ and $\mathcal{L}$ \cite{TorronKos13}. It is sufficient to follow the expectation values $E=\langle\hat{H}\rangle, L=\langle\hat{L}\rangle, C=\langle\hat{C}\rangle$ of the above operators, which evolve according to
\begin{eqnarray}
\label{E}\dot{E}&=&\left(\frac{\dot{\omega}}{\omega}+\gamma_a\omega^2\right)(E-L),\\
\label{L}\dot{L}&=&\left(-\frac{\dot{\omega}}{\omega}+\gamma_a\omega^2\right)E+\left[\frac{\dot{\omega}}{\omega}-(4\gamma_p+\gamma_a)\omega^2\right]L\nonumber\\
-2\omega C,\\
\label{C}\dot{C}&=&2\omega L+\left(\frac{\dot{\omega}}{\omega}-4\gamma_p\omega^2\right)C.
\end{eqnarray}
Note that the above equations are equivalent to Eq. (16) in \cite{TorronKos13}. In appendix \ref{app:stochastic} we show how they can be obtained using stochastic calculus, as an alternative to the open quantum systems formalism used here.

In this paper we concentrate on the evolution along the expansion phase of the cycle, where the frequency of the confining harmonic potential decreases from $\omega(0)=\omega_h$ to $\omega(T)=\omega_c$ at the final time $t=T$, so the working medium is actually expanded. The initial conditions are
\begin{equation}
\label{in_cond}
E(0)=E_h,\quad L(0)=C(0)=0,
\end{equation}
where $E_h$ is the initial energy while $L(0)=0$ corresponds to equipartition and $C(0)=0$ to the absence of correlations (ensemble in thermal equilibrium). Our goal is to find the frequency profile $\omega(t)$ which minimizes the final energy $E(T)=E_c$, maximizing thus the performance of the engine \cite{Salamon09}. The von Neumann entropy of the system is a monotonically increasing function of the following quantity, called the Casimir companion \cite{Boldt13}
\begin{equation}
\label{Casimir}
X=\frac{E^2-L^2-C^2}{\hbar^2\omega^2}.
\end{equation}
From (\ref{E}), (\ref{L}) and (\ref{C}) we find
\begin{equation}
\label{Casimir_der}
\dot{X}=\frac{2}{\hbar^2}[\gamma_a(E-L)^2+4\gamma_p(L^2+C^2)].
\end{equation}
Observe that in the absence of noise $(\gamma_a=\gamma_p=0)$, $X$ is a constant of the motion (Casimir invariant). In this case, the minimum final energy is obtained for $L(T)=C(T)=0$ and $\omega(T)=\omega_c$, and is given by \cite{Salamon09}
\begin{equation}
\label{min_energy}
E_c=\frac{\omega_c}{\omega_h}E_h.
\end{equation}
In the presence of noise, it is $\dot{X}\geq 0$, so $E_c\geq\omega_cE_h/\omega_h$. We would like to find $\omega(t)$ such that $E(T)=E_c$ is minimized and the final conditions
\begin{equation}
\label{fi_cond}
L(T)=C(T)=0
\end{equation}
are satisfied. We can quantify the performance of the engine relative to the noiseless case using the following measure \cite{TorronKos13}
\begin{equation}
\label{delta}
\delta=\frac{\omega_hE_c}{\omega_cE_h}-1\geq 0.
\end{equation}
Note that $\delta$ quantifies the decrease in the heat extraction efficiency of the engine due to noise \cite{TorronKos13}. In the ideal, noiseless case it is $\delta=0$; in general, the smaller is $\delta$ the better is the efficiency of the engine.

In the recent work \cite{TorronKos13} the authors consider the frequency profile
\begin{equation}
\label{mu}
\omega_n(t)=\frac{\omega_h}{1-\mu_n\omega_ht},\quad \mu_n=\frac{-2\ln\left(\frac{\omega_h}{\omega_c}\right)}{\sqrt{4n^2\pi^2+\ln^2\left(\frac{\omega_h}{\omega_c}\right)}}
\end{equation}
$n=1,2,\ldots$. With this choice, the final condition (\ref{fi_cond}) is satisfied in the ideal case for $t=T_n$, where
\begin{equation}
\label{Tn}
T_n=\frac{\left(\frac{\omega_h}{\omega_c}-1\right)\sqrt{4n^2\pi^2+\ln^2\left(\frac{\omega_h}{\omega_c}\right)}}{2\omega_h\ln\left(\frac{\omega_h}{\omega_c}\right)}
\end{equation}
In the presence of noise, the application of the above control input results in some remaining final energy in the $L$ and $C$ modes. The dissipation of this parasitic energy into the cold bath can stop the operation of the engine \cite{TorronKos13}. In the present paper we will use optimal control to find the frequency profile that minimizes the final energy and thus $\delta$, while satisfying as closely as possible the final conditions (\ref{fi_cond}).

%\begin{equation}
%\label{frequencybc}
%\omega(t)=\left\{\begin{array}{cc} \omega_{0}, & t\leq0\\\omega_{T}, & t\geq T\end{array}\right.,
%\end{equation}

%%%%%%%%%%%%%%%%%%%%%%%%%%%%%%%%%%%%%%%%%%%%%%%%%%%%%%%%%%%%%%%%%%%%%%%%%%%%%%%%

\section{MINIMIZING THE EFFECT OF NOISE USING OPTIMAL CONTROL}

\label{sec:Optimal_solution}

In order to bring the previously defined problem in a more appropriate form for the application of optimization methods, we use the state variables introduced in \cite{Tsirlin11,Salamon12}, with an additional normalization
\begin{equation}
\label{new_var}
x_1=\frac{\omega_h^2}{\omega^2}\frac{E-L}{E_h},\quad x_2=\frac{E+L}{E_h},\quad x_3=\frac{\omega_h}{\omega}\frac{C}{E_h}.
\end{equation}
Equations (\ref{E}), (\ref{L}) and (\ref{C}) become
\begin{eqnarray}
\label{x1}\dot{x}_1&=&-2\gamma_pux_1+2\gamma_px_2+2x_3,\\
\label{x2}\dot{x}_2&=&2(\gamma_a+\gamma_p)u^2x_1-2\gamma_pux_2-2ux_3,\\
\label{x3}\dot{x}_3&=&-ux_1+x_2-4\gamma_pux_3,
\end{eqnarray}
where time and noise strengths are normalized as
\begin{equation}
\label{new_par}
t_{\mathrm{new}}=\omega_ht,\quad \gamma_{a,\mathrm{new}}=\omega_h\gamma_a,\quad \gamma_{p,\mathrm{new}}=\omega_h\gamma_p,
\end{equation}
while the new control variable
\begin{equation}
\label{u}
u(t)=\frac{\omega^2(t)}{\omega_h^2}
\end{equation}
satisfies
\begin{equation}
\label{u_conditions}
u(0)=1,\quad u(T)=\frac{\omega_c^2}{\omega_h^2},\quad \frac{\omega_c^2}{\omega_h^2}\leq u(t)\leq 1.
\end{equation}
The initial conditions (\ref{in_cond}) become
\begin{equation}
\label{init_cond}
x_1(0)=x_2(0)=1,\quad x_3(0)=0.
\end{equation}
The old variables can be expressed as
\begin{equation}
\label{old_var}
\frac{E}{E_h}=\frac{x_2+ux_1}{2},\; \frac{L}{E_h}=\frac{x_2-ux_1}{2},\; \frac{C}{E_h}=\sqrt{u}x_3.
\end{equation}
It's not hard to see that the objective to find the control $u(t)$ minimizing $E(T)$ under the final conditions (\ref{fi_cond}) is equivalent to minimize $x_2(T)$ under the final conditions
\begin{equation}
\label{final_cond}
x_2(T)-u(T)x_1(T)=0,\quad x_3(T)=0.
\end{equation}

The main advantage of using the new variables (\ref{new_var}) is that $\dot{\omega}$ does not appear in the new state equations, so the control $u(t)$ in (\ref{u}) can contain even jump discontinuities \cite{Tsirlin11,Salamon12}. Indeed, it can be shown that in the absence of noise the optimal control achieving the minimum value of the final energy (\ref{min_energy}) in minimum time has the bang-bang form, jumping between and waiting at the extreme values of $u$ \cite{Salamon09,Tsirlin11,Salamon12}. When noise is present ($\gamma_a>0$ or $\gamma_p>0$), the control appears quadratically in (\ref{x2}). In this case, optimal control theory \cite{Pontryagin} implies that the optimal control may contain, additionally to the bang segments, time intervals where $u(t)$ is continuous. Now, an important observation is that no matter what variables we use, we face a two-point boundary value problem, as can be seen from conditions (\ref{u_conditions}),(\ref{init_cond}), and (\ref{final_cond}), in a three-dimensional nonlinear control system. For such problems it is in general difficult to find an analytical solution, and we have to rely on numerical optimization. We choose the Legendre pseudospectral method for optimal control \cite{Ross03}. This method, extensively used for trajectory optimization in aerospace applications \cite{Bedrossian12}, has been proven successful for pulse design in open quantum systems, in the context of Nuclear Magnetic Resonance spectroscopy \cite{LiJCP09}. The idea behind the method is to convert a continuous-time optimal control problem to a discrete nonlinear programming problem, which can be solved by many well-developed computational algorithms.

In order to apply the proposed method, it is first necessary to transform the problem from the time interval $t\in [0,T]$ to $\tau\in[-1,1]$, using the transformation $\tau=(2t-T)/T$. In a redundant use of notation, we make this transition and reuse the same time variable $t$. The next step is to approximate the states $x_r(t), r=1,2,3$ and the control $u(t)$ by the $N$th order interpolating polynomials $I_Nx_r(t), I_Nu(t)$ in the Lagrange polynomial basis $\ell_i(t)$
\begin{eqnarray}
\label{x_disc}x_r(t)  &  \simeq  & I_Nx_r(t) = \sum_{i=0}^N x_{ri}\ell_i(t),\\
\label{u_disc}u(t)  &  \simeq  & I_Nu(t) =  \sum_{i=0}^N u_i\ell_i(t).
\end{eqnarray}
By using the $N+1$ Legendre-Gauus-Lobatto (LGL) interpolation nodes, the error in the above approximations is close to minimum \cite{Canuto06}. The LGL grid is comprised by the endpoints $t_0=-1, t_N=1$ and the $N-1$ roots of the derivative of the $N$th order Legendre polynomial. From the property $\ell_i(t_j)=\delta_{ij}$ of the Lagrange polynomials we have $I_Nx_r(t_j)=x_{rj}=x_r(t_j)$, $I_Nu(t_j)=u_j=u(t_j)$ for $j=0\ldots N$.

The Legendre pseudospectral method is a collocation method where the dynamics is enforced at the LGL nodes. The derivative of $I_Nx_r(t)$ at the LGL node $t_k$
is given by \cite{LiJCP09}
\begin{equation}
\frac{d}{dt}I_Nx_r(t_k)=\sum_{i=0}^N x_{ri}\dot{\ell}_i(t_k)=\sum_{i=0}^ND_{ki}x_{ri},
\end{equation}
where $D_{ki}$ are elements of the constant $(N+1)\times(N+1)$ differentiation matrix $D$ defined by \cite{Canuto06}
\begin{equation}\label{eq:D}
D_{ki} = \left\{
\begin{array}{cl}
    \frac{L_N(t_k)}{L_N(t_i)}\frac{1}{t_k-t_i} & k \neq i \\  & \\
    -\frac{N(N+1)}{4} & k=i=0 \\ & \\
    \frac{N(N+1)}{4} & k=i=N \\ & \\
    0 & \textrm{otherwise}.
\end{array}
\right.
\end{equation}
Equations (\ref{x1}), (\ref{x2}), and (\ref{x3}) take the discrete form
\begin{eqnarray}
\label{x1_d}\frac{2}{T}\sum_{i=0}^ND_{ki}x_{1i} & = & -2\gamma_pu_kx_{1k}+2\gamma_px_{2k}+2x_{3k},\\
\label{x2_d}\frac{2}{T}\sum_{i=0}^ND_{ki}x_{2i} & = & 2(\gamma_a+\gamma_p)u^2_kx_{1k}-2\gamma_pu_kx_{2k}\nonumber\\
-2u_kx_{3k},\\
\label{x3_d}\frac{2}{T}\sum_{i=0}^ND_{ki}x_{3i} & = & -u_kx_{1k}+x_{2k}-4\gamma_pu_kx_{3k},
\end{eqnarray}
where $k=0\ldots N$ and the factor $2/T$ on the left hand sides comes from the initial change in the time variable.
The initial conditions (\ref{init_cond}) become
\begin{equation}
\label{init_cond_d}
x_{10}=x_{20}=1,\quad x_{30}=0,
\end{equation}
and, analogously, the final conditions (\ref{final_cond})
\begin{equation}
\label{final_cond_d}
x_{2N}-u_Nx_{1N}=0,\quad x_{3N}=0.
\end{equation}
The control constraints (\ref{u_conditions}) take the form
\begin{equation}
\label{u_conditions_d}
u_0=1\frac{}{},\quad u_N=\frac{\omega_c^2}{\omega_h^2},\quad \frac{\omega_c^2}{\omega_h^2}\leq u_k\leq 1.
\end{equation}
The objective is to find $x_{rk}$ and $u_k$, where $r=1,2,3$ and $k=0\ldots N$, such that $x_{2N}$ is minimized and the above conditions are satisfied.

Observe that the original continuous-time optimal control problem has been transformed to a discrete nonlinear programming problem, which can be solved by existent software packages. In this article we use AMPL (A Mathematical Programming Language) \cite{AMPL} with MINOS 5.5 solver. After calculating the optimal $u_k$, we interpolate them to obtain a continuous control $u(t)$. We subsequently apply this control to system equations (\ref{x1})-(\ref{x3}) using MATLAB function ode45, and record the resultant efficiency. Note that all the results of the next section are obtained using $N=69$.

%\begin{equation}
%\mathbf{f}=\left(
%\begin{array}
%[c]{c}
%x_{2}\\
%1/x_{1}^{2}
%\end{array}
%\right)  ,\,\,\mathbf{g}=\left(
%\begin{array}
%[c]{c}
%0\\
%-x_{1}
%\end{array}
%\right),\nonumber
%\end{equation}

%\begin{figure}[t]
%\centering
%\includegraphics[width=0.9\linewidth]{Comparison}\caption{(Color online) One- and multi-switchings trajectories. Solid line corresponds to $u=1$, dashed line to $u=-1$.}
%\label{fig:comparison}
%\end{figure}

\section{RESULTS AND DISCUSSION}
\label{sec:comparison}

\begin{figure}[t]
 \centering
		\begin{tabular}{c}
     	\subfigure{
	            \label{fig:compdeph1}
	            \includegraphics[width=0.7\linewidth]{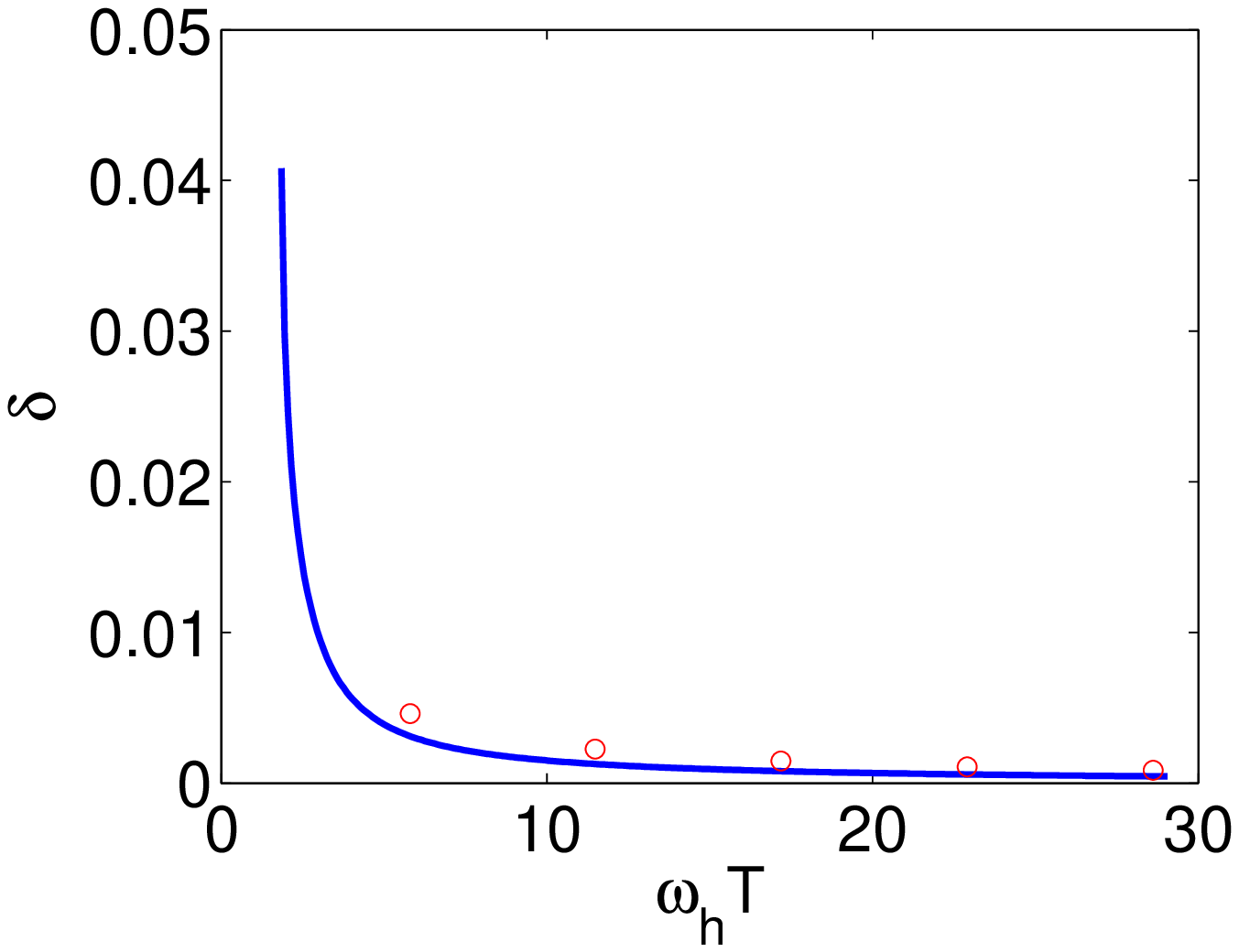}} \\
	    \subfigure{
	            \label{fig:compdeph2}
	            \includegraphics[width=0.7\linewidth]{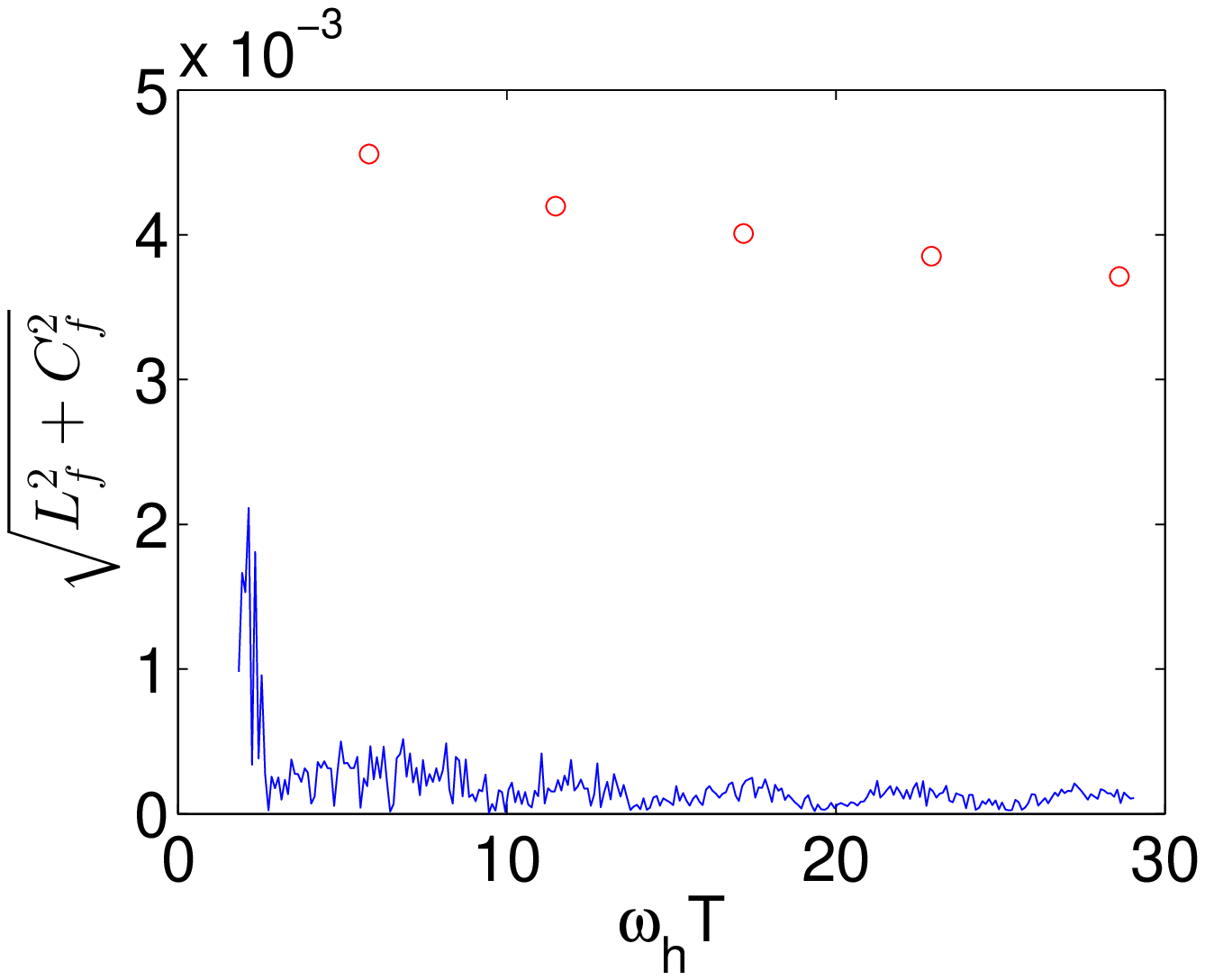}}
		\end{tabular}
\caption{(Color online) (a) For $\gamma_a=0, \omega_h\gamma_p=0.01$ (phase damping) and $\omega_c/\omega_h=1/3$ we plot the optimal parameter $\delta$ (blue solid line), as well as the same quantity for the frequency profile (\ref{mu}) used in \cite{TorronKos13} and for durations $T_n, n=1,2,3,4,5$ from (\ref{Tn}) (red circles). Observe that the optimized $\delta$ is lower, corresponding to a higher efficiency. Note that there is a minimum necessary time for a feasible solution of the optimization problem, $\omega_hT=1.85$. For large $T$ parameter $\delta$ approaches zero, the ideal value corresponding to the noiseless case. The reason is that for phase damping and long enough durations, the evolution of the system can take place close to a noise-free path. (b) For both cases we display a measure of the undesirable remaining energy in the $L, C$ modes. Observe that this parasitic energy is lower in the optimized case (blue solid line).}
\label{fig:compdeph}
\end{figure}

Using the optimization method described in the previous section, we can find the controls maximizing the efficiency of the engine for various values of the parameters. In Fig. \ref{fig:compdeph1} we plot the numerically obtained optimal quantity $\delta$ (\ref{delta}) versus the normalized duration $\omega_hT$, in the case of pure dephasing $\gamma_a=0$, $\omega_h\gamma_p=0.01$ and for the ratio $\omega_c/\omega_h=1/3$ (blue solid line). For comparison, we also display the same quantity for the frequency profile (\ref{mu}) used in \cite{TorronKos13} and for durations $T_n, n=1,2,3,4,5$ from (\ref{Tn}) (red circles). Obviously, a smaller value of $\delta$ (and thus a better engine performance) corresponds to the optimized case. Also observe that there is a minimum time such that the optimization problem defined in section \ref{sec:Optimal_solution} has a feasible solution. For the specific parameter values, this minimum time is $\omega_hT=1.85$. This time is larger than the minimum necessary time to obtain the ideal performance $\delta=0$ in the absence of noise, which can be found in \cite{Salamon09} and is $\omega_hT=1.79$ for our example. In Fig. \ref{fig:compdeph2} we plot for both cases the quantity $\sqrt{L_f^2+C_f^2}=\sqrt{L^2(T)+C^2(T)}$, which is a measure of the undesirable remaining energy in the $L, C$ modes. We observe that this parasitic energy is lower in the optimized case (blue solid line). Although one may expect this quantity to be zero for the optimized case, there is a remaining value due to the discretization and the interpolation that we use to obtain the continuous $u(t)$, as explained in the last paragraph of the previous section.

\begin{figure}[t]
 \centering
		\begin{tabular}{cc}
     	\subfigure[$\ $$\omega_hT=1.85$]{
	            \label{fig:control11}
	            \includegraphics[width=.45\linewidth]{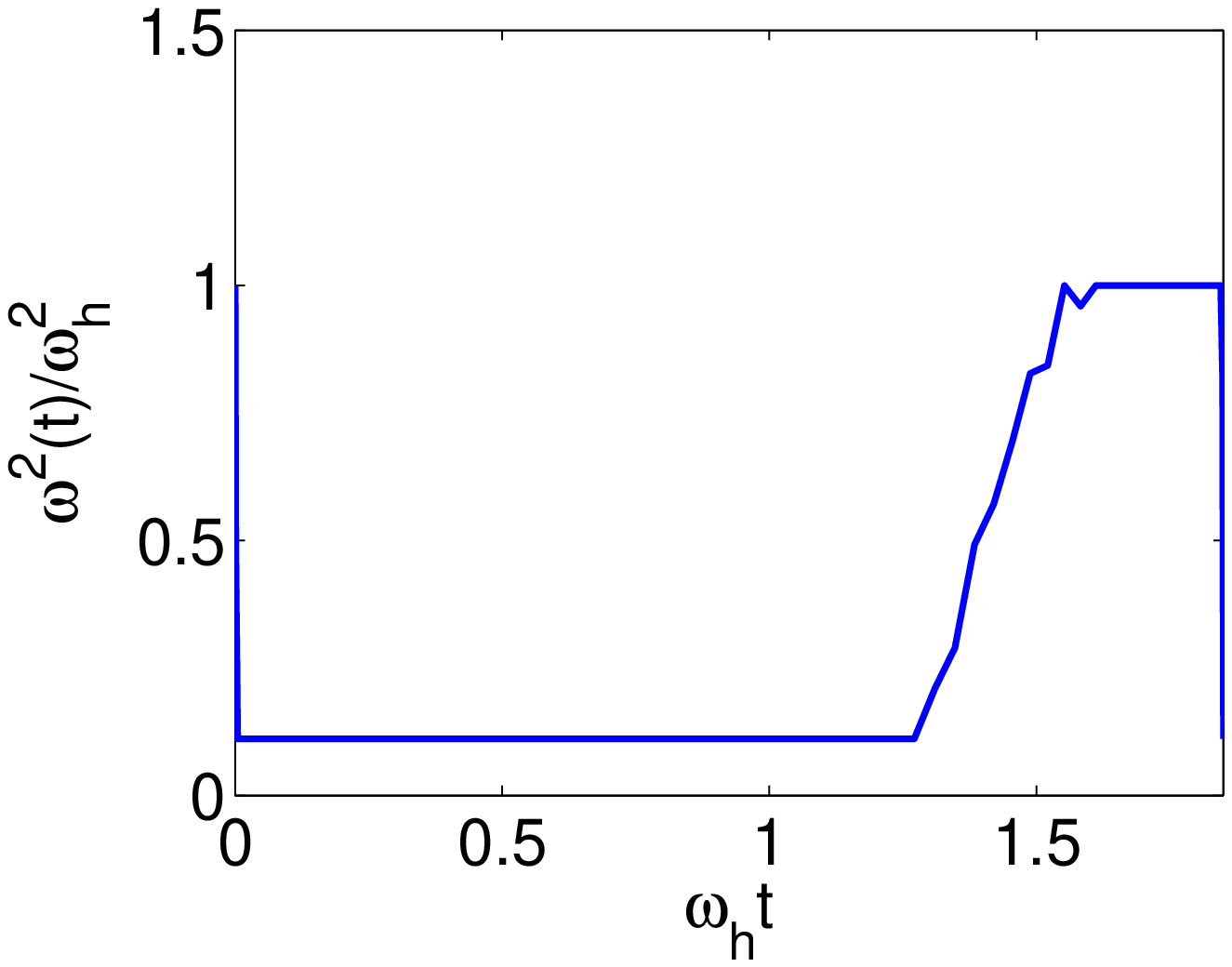}} &
       \subfigure[$\ $$\omega_hT=2$]{
	            \label{fig:control12}
	            \includegraphics[width=.45\linewidth]{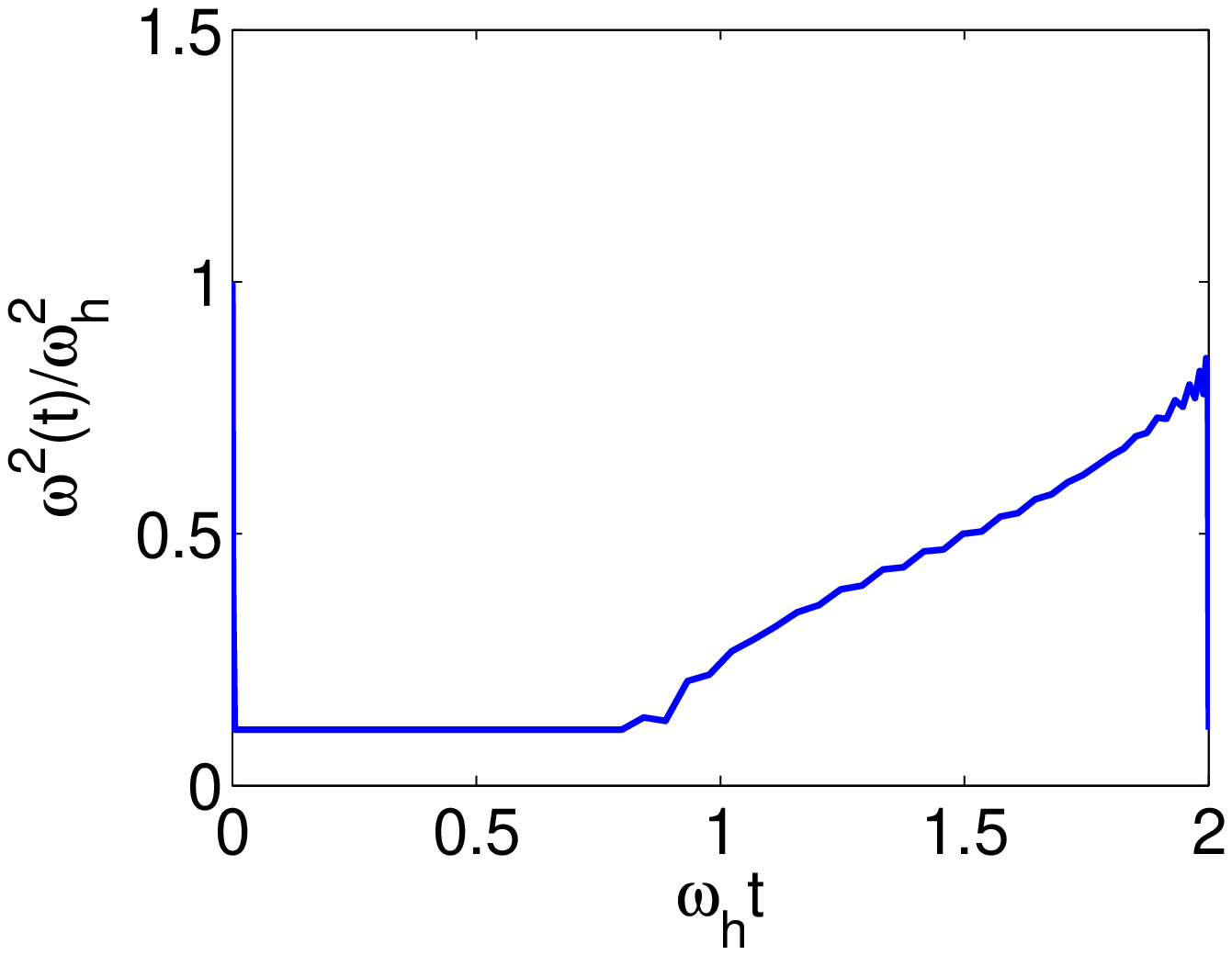}} \\
	    \subfigure[$\ $$\omega_hT=5$]{
	            \label{fig:control13}
	            \includegraphics[width=.45\linewidth]{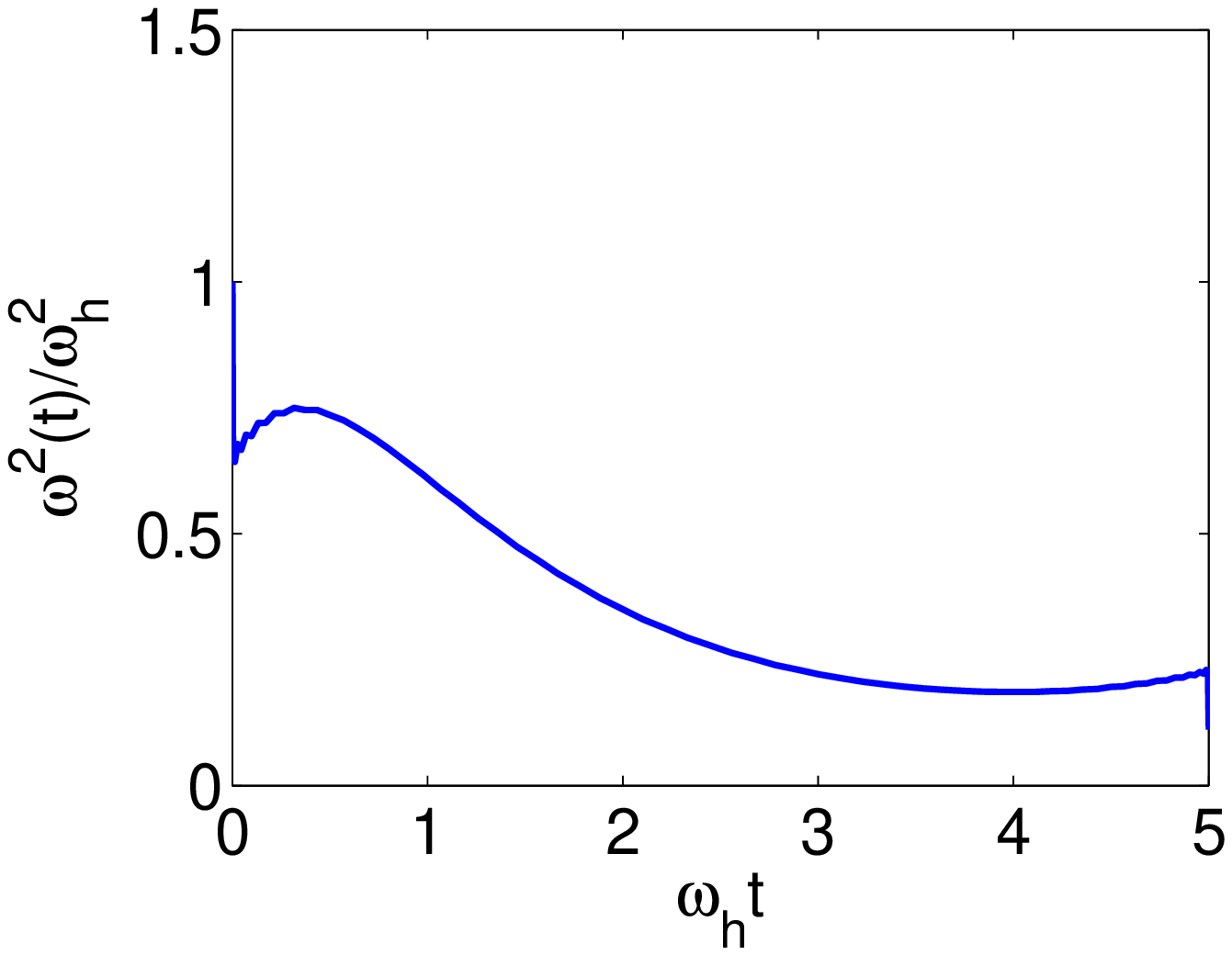}} &
       \subfigure[$\ $$\omega_hT=29$]{
	            \label{fig:control14}
	            \includegraphics[width=.45\linewidth]{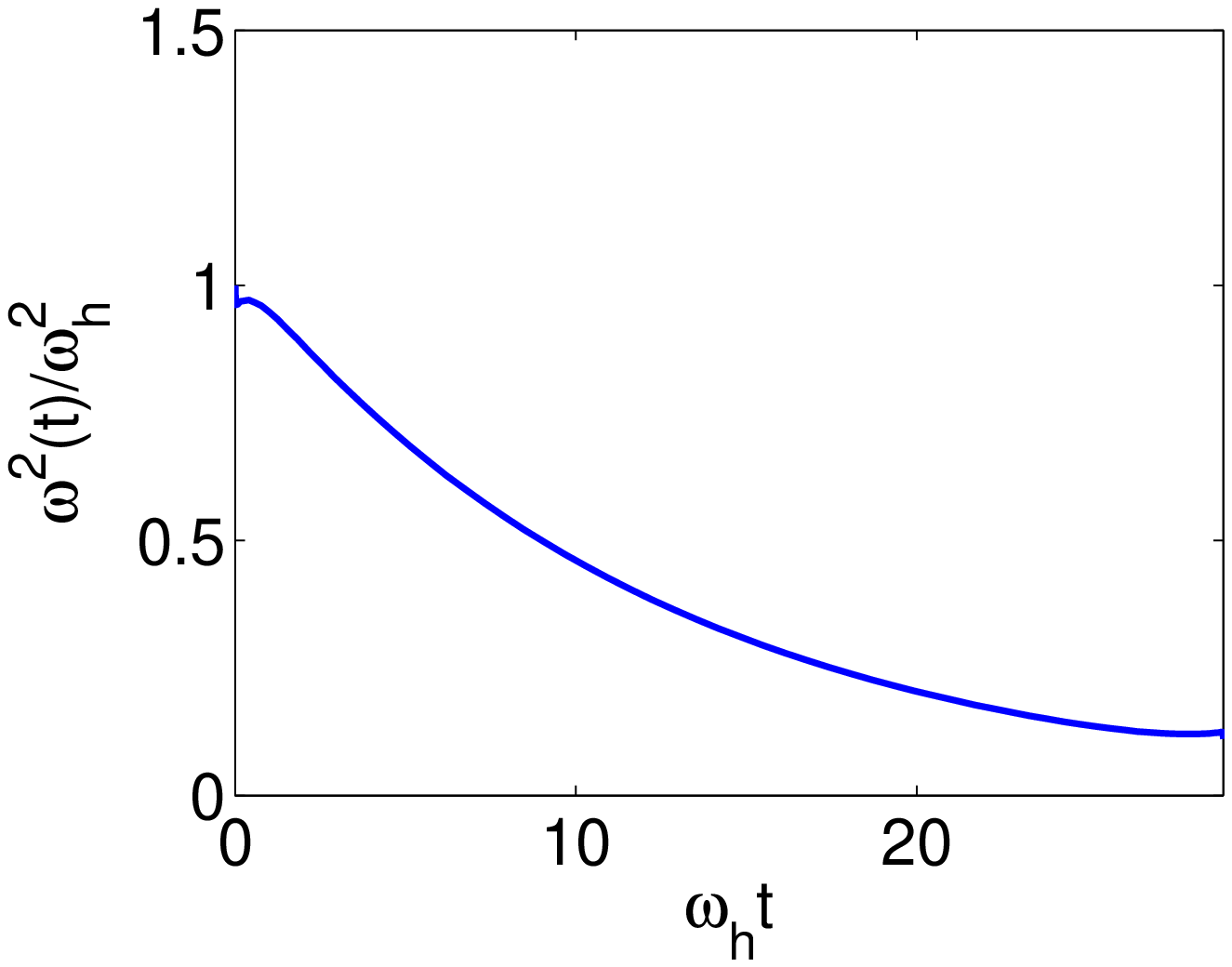}}
		\end{tabular}
\caption{(Color online) For $\gamma_a=0, \omega_h\gamma_p=0.01$ and $\omega_c/\omega_h=1/3$ we plot the numerically obtained optimal frequency profile for various values of the normalized time. Observe that for shorter available times the optimal frequency takes values on the boundary of the allowed region (\ref{u_conditions}), while for larger times it is smoother.}
\label{fig:controls1}
\end{figure}

Note that in the case of pure dephasing $\gamma_a=0, \gamma_p>0$ the Casimir companion $X$ is conserved for evolution along the direction $(E,0,0)$, as can be easily seen from (\ref{Casimir_der}). For large enough $T$, the system can evolve closely to this direction following an almost noise-free path, as explained in appendix \ref{app:bound}. In the limiting case $T\rightarrow\infty$, the maximum efficiency of the ideal case $\delta=0$ can be obtained despite the presence of dephasing, as we show in appendix \ref{app:bound} and is depicted in Fig. \ref{fig:compdeph1}.  The situation is reminiscent of STIRAP in a $\Lambda$-type atom, where perfect population transfer is achieved between two ground states coupled through a lossy excited state, which is actually never populated in the adiabatic (long time) limit. The attainment of ideal performance in the presence of dephasing can be attributed to the existence of a path along the noise-free subspace $(E,0,0)$ connecting the initial and final states \cite{Yuan12}. Note that this result seems to be in contrast with the finite limiting value for $\delta$ obtained in \cite{TorronKos13}, Eq. (32) there.

In Fig. \ref{fig:controls1} we plot the optimal input $u(t)=\omega^2(t)/\omega^2_h$ for various values of the duration $T$ and for the same parameters as before. Observe that for short $T$ the optimal control contains bang segments, where it takes values on the boundaries, while for larger $T$ it is smoother. This control shape can be understood considering that for short durations the major efficiency bottleneck is not the noise but the limited available time. It is also consistent with the optimal control form expected from the application of Pontryagin's maximum principle \cite{Pontryagin} to our system, where the bounded control $u$ enters quadratically the state equation (\ref{x2}) when noise is present.

\begin{figure}[t]
 \centering
		\begin{tabular}{c}
     	\subfigure{
	            \label{fig:complong1}
	            \includegraphics[width=0.7\linewidth]{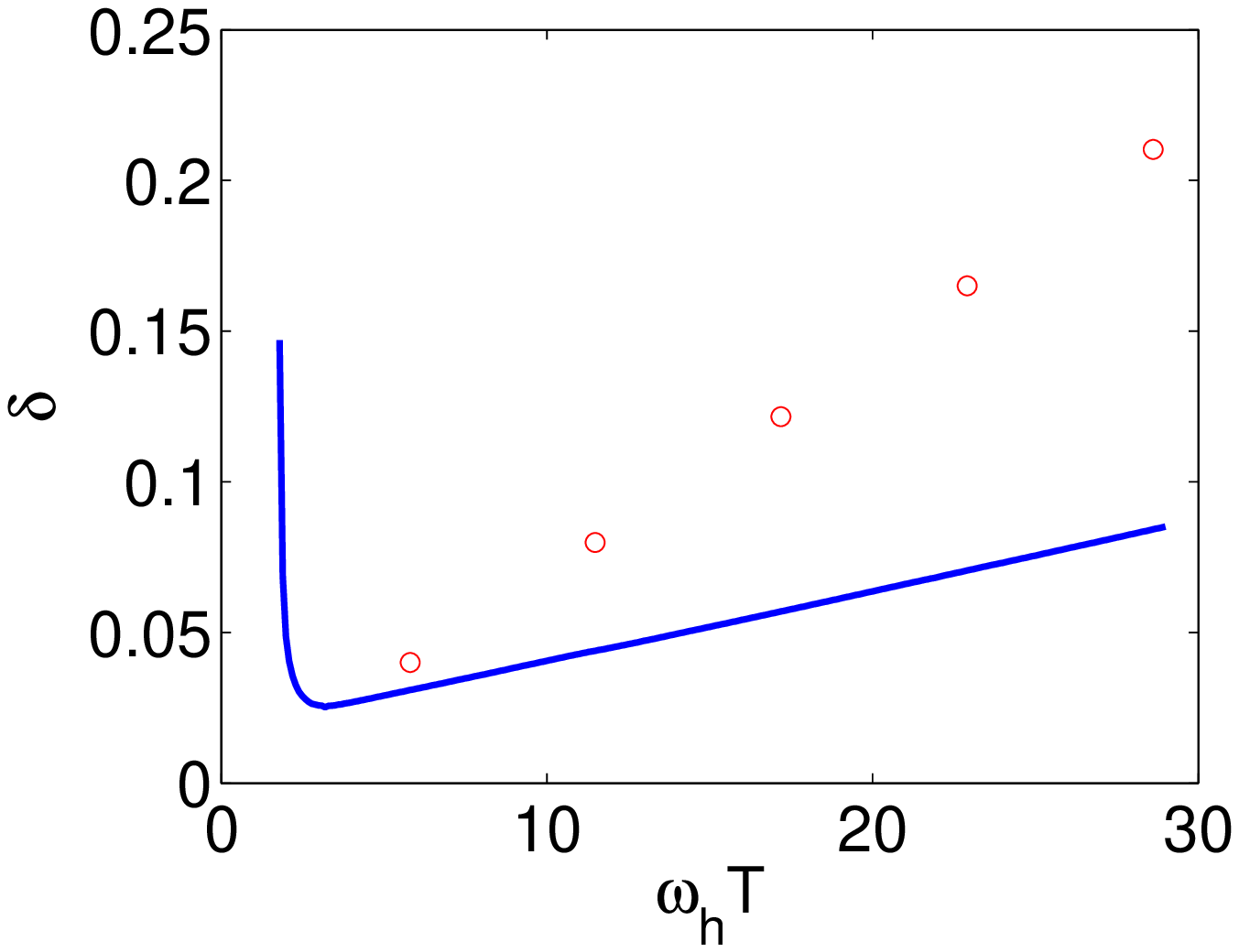}} \\
	    \subfigure{
	            \label{fig:complong2}
	            \includegraphics[width=0.7\linewidth]{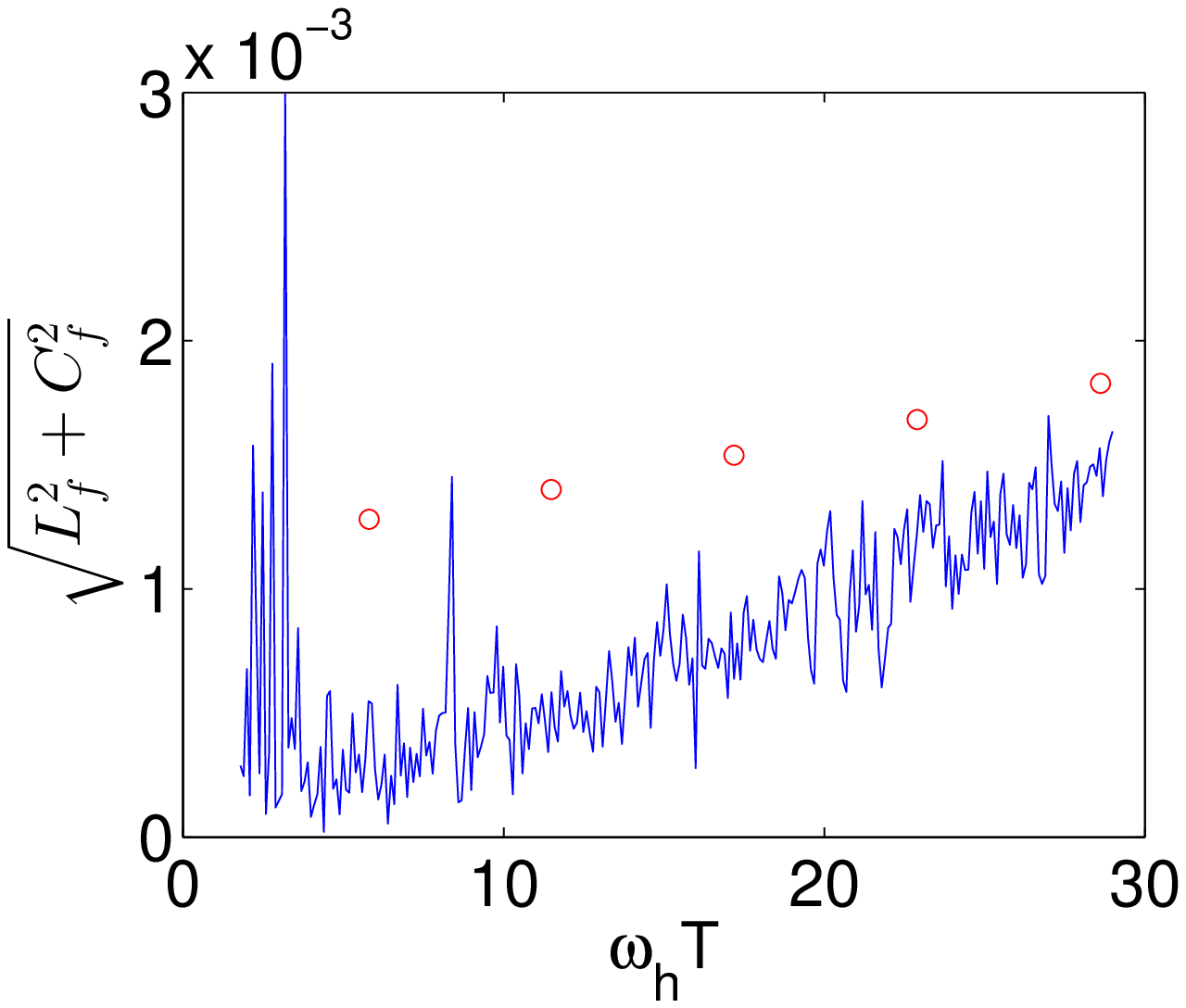}}
		\end{tabular}
\caption{(Color online) (a) For $\omega_h\gamma_a=0.02, \gamma_p=0$ (amplitude noise) and $\omega_c/\omega_h=1/3$ we plot the optimal parameter $\delta$ (blue solid line), as well as the same quantity for the frequency profile (\ref{mu}) used in \cite{TorronKos13} and for durations $T_n, n=1,2,3,4,5$ from (\ref{Tn}) (red circles). Again, the optimized $\delta$ is lower, corresponding to a higher efficiency, and there is a minimum necessary time for a feasible solution of the optimization problem, $\omega_hT=1.89$. For small $T$ the quantity $\delta$ increases due to the limited available time, while for large $T$ increases since for amplitude noise the evolution does not take place along a noise-free path. As a result, there is an intermediate time $T$ where $\delta$ is minimized. (b) The parasitic energy is again lower in the optimized case (blue solid line).}
\label{fig:complong}
\end{figure}

\begin{figure}[t]
 \centering
		\begin{tabular}{cc}
     	\subfigure[$\ $$\omega_hT=1.89$]{
	            \label{fig:control21}
	            \includegraphics[width=.45\linewidth]{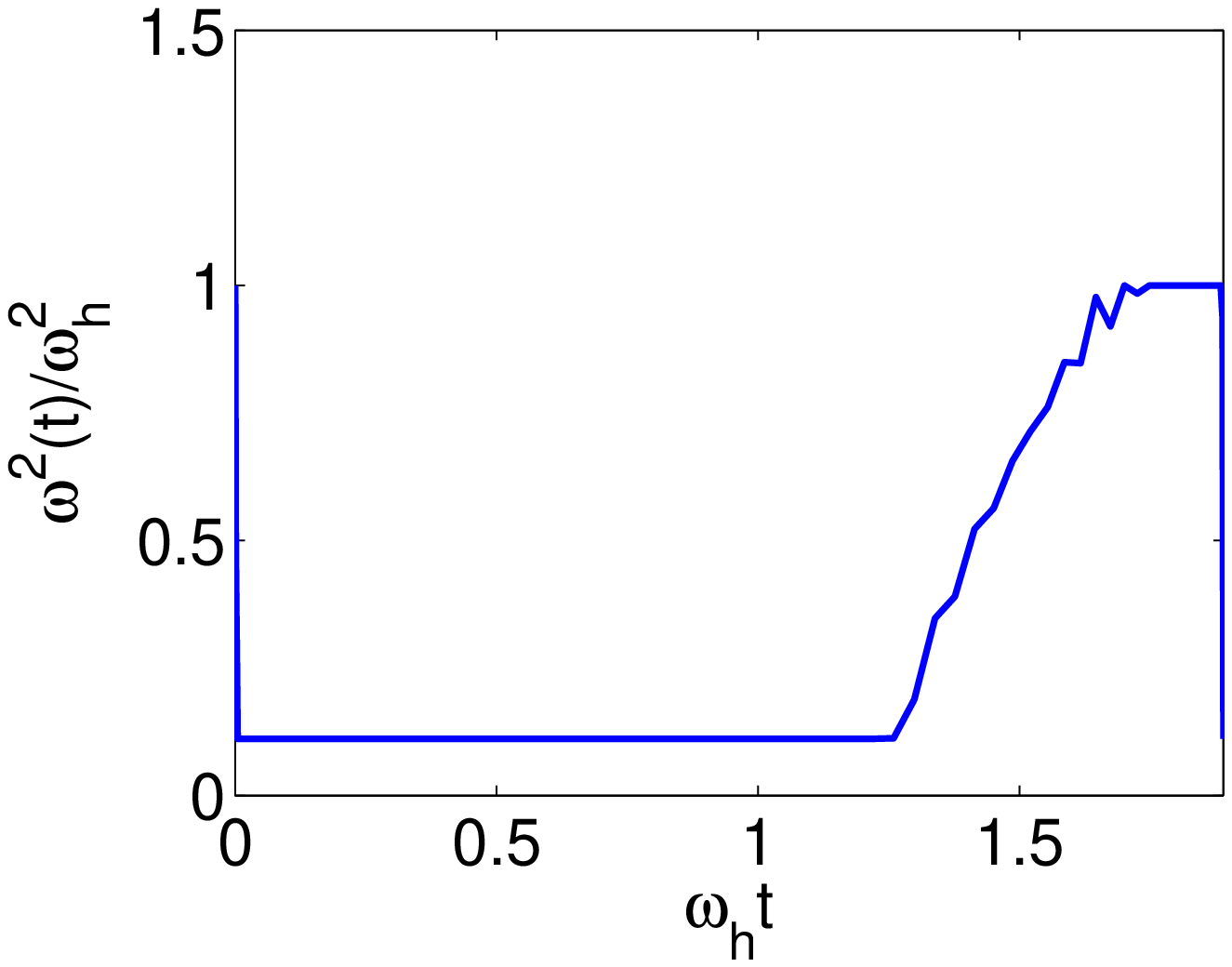}} &
       \subfigure[$\ $$\omega_hT=2$]{
	            \label{fig:control22}
	            \includegraphics[width=.45\linewidth]{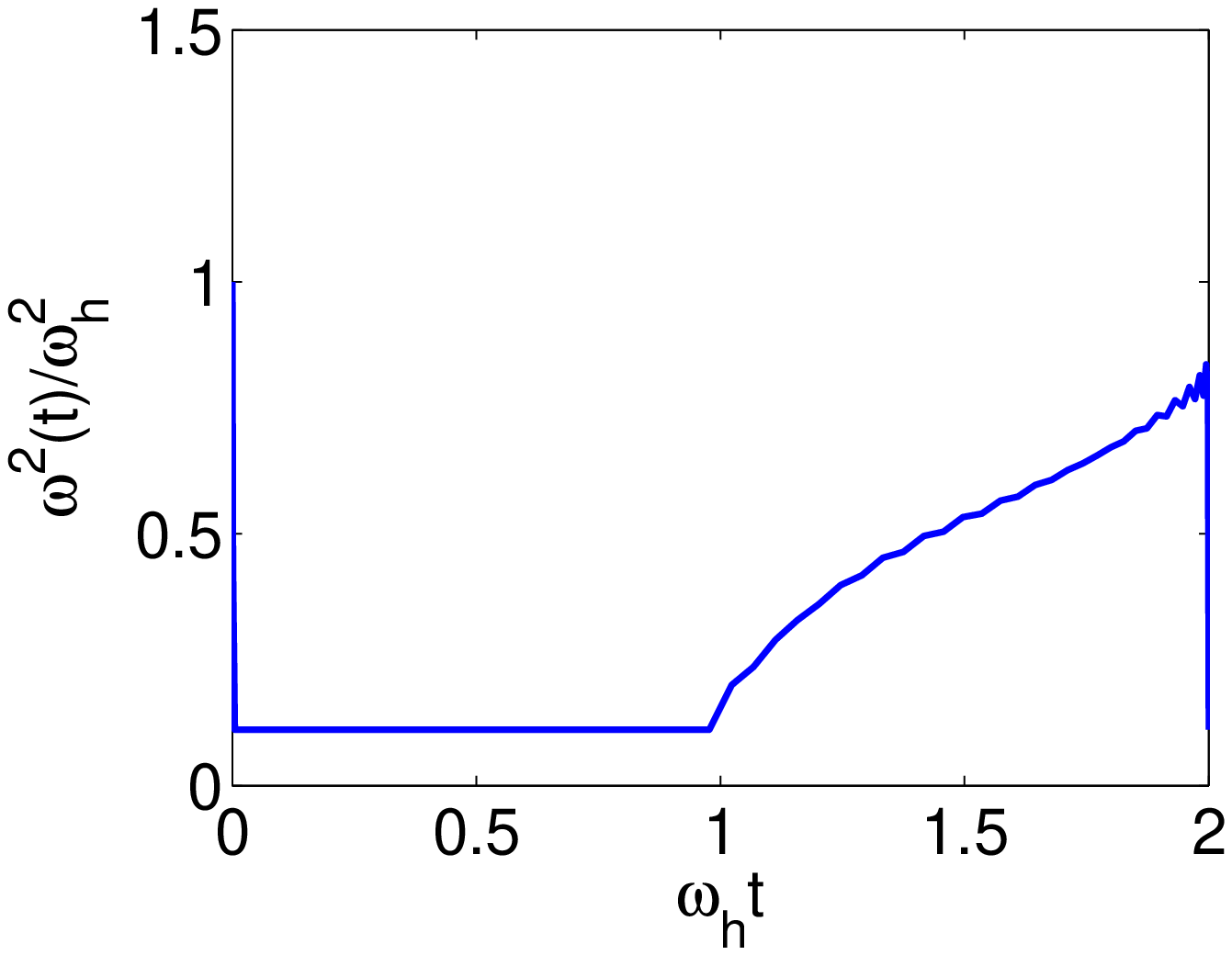}} \\
	    \subfigure[$\ $$\omega_hT=5$]{
	            \label{fig:control23}
	            \includegraphics[width=.45\linewidth]{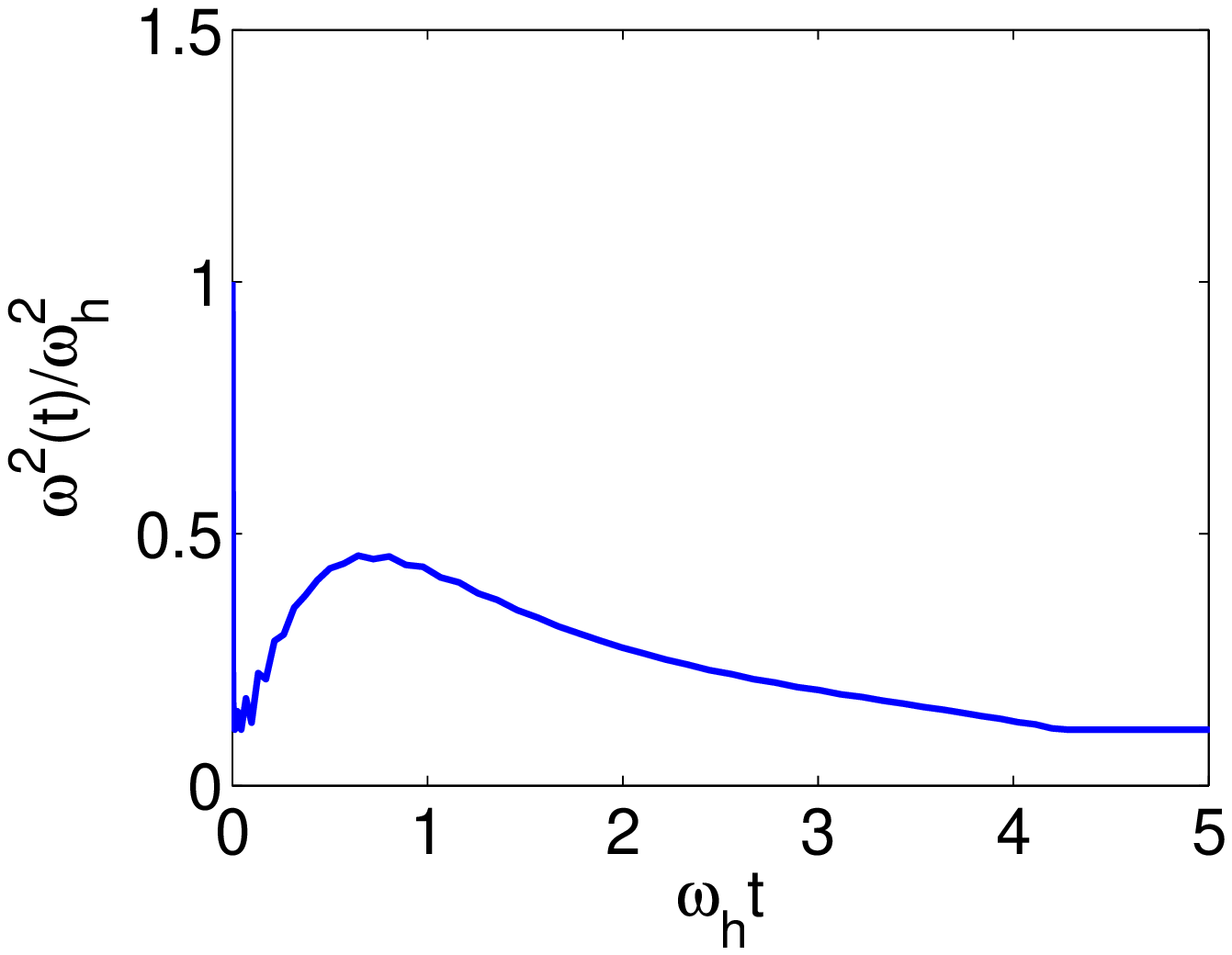}} &
       \subfigure[$\ $$\omega_hT=29$]{
	            \label{fig:control24}
	            \includegraphics[width=.45\linewidth]{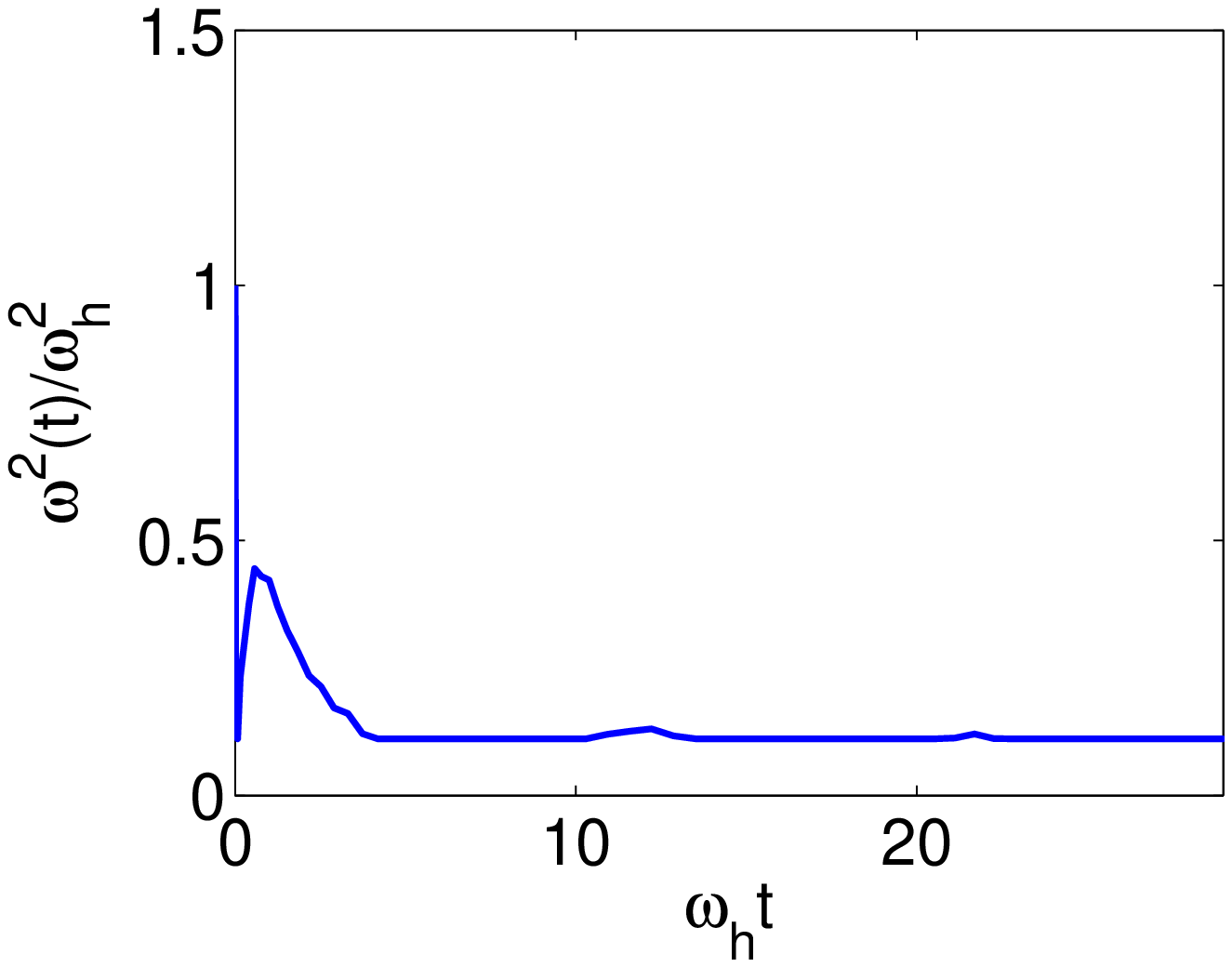}}
		\end{tabular}
\caption{(Color online) For $\omega_h\gamma_a=0.02, \gamma_p=0$ and $\omega_c/\omega_h=1/3$ we plot the numerically obtained optimal frequency profile for various values of the normalized time. Observe that for shorter available times the optimal shape is similar to the corresponding cases of Fig. \ref{fig:controls1}, while for larger times where relaxation dominates it is different.}
\label{fig:controls2}
\end{figure}

In Fig. \ref{fig:complong1} we plot the optimal $\delta$ versus the normalized time $\omega_hT$ for $\omega_h\gamma_a=0.02$, $\gamma_p=0$ and for the same ratio $\omega_c/\omega_h=1/3$ as before (blue solid line). Again, there is a minimum necessary time such that the optimization problem has a feasible solution, which is $\omega_hT=1.89$. For short $T$ close to the minimum, $\delta$ is large. As $T$ increases, $\delta$ decreases and attains a minimum value. After that, $\delta$ increases with increasing $T$. This dependence of $\delta$ on $T$ can be understood in the following way. For short $T$, the small available time limits the performance, while for large $T$ the efficiency is degraded by the noise. Note that for $\gamma_a>0, \gamma_p=0$, a noise-free path would require $E=L$, as we can observe from (\ref{Casimir_der}). But this is not the case for the transfer that we examine, since $E>0$ and $L=0$ at the initial and final times. As a consequence, for this type of noise the performance is reduced for larger times. In the same figure, we also plot $\delta$ for the input (\ref{mu}) used in \cite{TorronKos13} and for durations $T_n, n=1,2,3,4,5$ from (\ref{Tn}) (red circles). Again, these values of $\delta$ are larger than those corresponding to the optimized case. In Fig. \ref{fig:complong2} we display the quantity $\sqrt{L_f^2+C_f^2}$ for the same inputs, and we observe that the remaining parasitic energy is lower for the optimized case.

In Fig. \ref{fig:controls2} we plot the optimal input $u(t)=\omega^2(t)/\omega^2_h$ for various values of the duration $T$ and for the same parameters used in Fig. \ref{fig:complong}. Observe that for short $T$, where the main efficiency limitation is the small available time, the shape of the optimal control is similar to the previous case of pure dephasing, compare Figs. \ref{fig:control11}, \ref{fig:control12} with Figs. \ref{fig:control21}, \ref{fig:control22}. For longer times, where the noise plays the major role, the optimal shape is different for the different noise mechanisms, compare Figs. \ref{fig:control13}, \ref{fig:control14} with Figs. \ref{fig:control23}, \ref{fig:control24}.

When both phase and amplitude noise are present, the dependence of $\delta$ on $T$ is similar to that shown in Fig. \ref{fig:complong1}, i.e. there is an optimal $T$ where $\delta$ is minimized, as explained in \cite{TorronKos13}. The location of the minimum depends on the ratio $\gamma_p/\gamma_a$. In general, for larger values of this ratio the minimum is shifted towards higher $T$, while for lower values it is shifted towards smaller $T$.

%%%%%%%%%%%%%%%%%%%%%%%%%%%%%%%%%%%%%%%%%%%%%%%%%%%%%%%%%%%%%%%%%%%%%%%%%%%%%%%%
\section{CONCLUSION AND FUTURE WORK}

\label{sec:conclusion}

In this paper, we applied optimal control to maximize the performance of a quantum heat engine executing the Otto cycle in the presence of external noise. We have shown numerically that there is an improvement in the engine efficiency compared to that obtained with the input used in \cite{TorronKos13}, for both amplitude and phase noise. In the case of phase damping we have additionally proved that the ideal performance of a noiseless engine can be retrieved in the adiabatic limit $T\rightarrow\infty$. These results can find application in the quest for absolute zero, i.e. the design of quantum refrigerators that can cool a quantum system to the lowest possible temperature. They are also directly applicable to the optimal control of a collection of classical harmonic oscillators sharing the same time-dependent frequency \cite{Hoffmann13} and subjected to similar noise mechanisms.

An interesting extension of the present work is to examine how the results are modified if we allow the control input $u(t)=\omega^2(t)/\omega^2_h$ to take negative values, corresponding to a repulsive parabolic potential for some finite time interval. Since the control set is augmented, a better efficiency is expected. Another important question is trying to prove rigorously the existence of a feasible solution for the optimal control problem defined in section \ref{sec:Optimal_solution}. Finally, note that the methodology employed in this work can be used for the optimization of other interesting thermodynamic processes.

%%%%%%%%%%%%%%%%%%%%%%%%%%%%%%%%%%%%%%%%%%%%%%%%%%%%%%%%%%%%%%%%%%%%%%%%%%%%%%%%

%\section*{ACKNOWLEDGEMENTS}

%The author would like to thank RR for her patience.

\appendix

\section{DERIVATION OF THE SYSTEM EQUATIONS USING STOCHASTIC CALCULUS}

\label{app:stochastic}

In this appendix we use stochastic calculus to derive Eqs. (\ref{E})-(\ref{C}), which are derived in the main text using a master equation formalism. Consider the following stochastic Hamiltonian $\hat{\mathcal{H}}_s$, where $n_a(t), n_p(t)$ are independent, zero mean gaussian white noises
\begin{eqnarray}
\label{Sto_Hamiltonian}
\hat{\mathcal{H}}_s&=&\frac{\hat{p}^2}{2m}+\frac{m\omega^2(t)}{2}(1+n_a(t))\hat{q}^2+n_p(t)\hat{H}\nonumber\\
                   &=&\hat{H}+\frac{m\omega^2(t)}{2}\hat{q}^2n_a(t)+\hat{H}n_p(t).
\end{eqnarray}
Obviously $n_a(t)$ represents fluctuations in the stiffness of the oscillator, while $n_p(t)$ corresponds to phase damping. The evolution of an operator $\hat{A}$ \emph{before averaging} is given by the following stochastic differential equation in the Stratonovich sense
\begin{equation}
\label{Sto}
\frac{d\hat{A}}{dt}=\frac{i}{\hbar}[\hat{\mathcal{H}}_s,\hat{A}]+\frac{\partial\hat{A}}{\partial t}.
\end{equation}
Observe that the operators $\hat{H},\hat{L},\hat{C}$ are functions of the triplet $(\hat{p},\hat{q},t)$. The stochastic equation for $\hat{p}$ is (in the following we use the Stratonovich differential $\dbar$ and the Wiener processes $w_a,w_p$ corresponding to the noises $n_a=\dbar w_a/dt,n_p=\dbar w_p/dt$)
\begin{eqnarray}
\label{p}
\dbar \hat{p}&=&\frac{i}{\hbar}[\hat{H}dt+\frac{m\omega^2\hat{q}^2}{2}\dbar w_a+\hat{H}\dbar w_p,\hat{p}]\nonumber\\
             &=&-m\omega^2\hat{q}(dt+\dbar w_a+\dbar w_p),
\end{eqnarray}
while for $\hat{q}$ we find
\begin{equation}
\label{q}
\dbar \hat{q}=\frac{\hat{p}}{m}(dt+\dbar w_p).
\end{equation}
For the Stratonovich calculus, widely used in Physics, the usual product rule $\dbar (a\cdot b)=\dbar a\cdot b+a\cdot\dbar b$ holds (but not for the It\={o} calculus used in finance). For the Hamiltonian $\hat{H}$, given in (\ref{Hamiltonian}) as a function of $(\hat{p},\hat{q},t)$, we find
\begin{equation}
\label{Hamil}
\dbar \hat{H}=\frac{1}{2m}(\hat{p}\dbar\hat{p}+\dbar\hat{p}\hat{p})+\frac{m\omega^2}{2}(\hat{q}\dbar\hat{q}+\dbar\hat{q}\hat{q})+m\omega\dot{\omega}\hat{q}^2dt.
\end{equation}
Using (\ref{p}), (\ref{q}) we end up with
\begin{equation}
\label{Hdiff}
\dbar \hat{H}=\frac{\dot{\omega}}{\omega}(\hat{H}-\hat{L})dt-\frac{\omega^2}{2}(\hat{p}\hat{q}+\hat{q}\hat{p})\dbar w_a.
\end{equation}
Note that Stratonovich calculus is partially anticipatory and the term multiplying the noise in (\ref{Hdiff}) is correlated with the noise. In order to find the average over the noise, it is easier to use the non-anticipatory It\={o} calculus. First we have to find the It\={o} differential $d\hat{H}$ corresponding to (\ref{Hdiff}). Following the rules described in \cite{BrockettStoCon}, we find
\begin{equation}
\label{HIto}
d\hat{H}=\left[\frac{\dot{\omega}}{\omega}(\hat{H}-\hat{L})+\gamma_a\omega^2(\hat{H}-\hat{L})\right]dt-\omega\hat{C}dw_a.
\end{equation}
Observe the extra term multiplying $dt$ and note that $dw_a\cdot dw_a=2\gamma_adt$ (Wiener process of appropriate strength), while the independence of $w_a,w_p$ has also been used in the derivation of the above equation. Now we can take the average over the noise, keeping the same symbols for the operators, and find the deterministic equation
\begin{equation}
\label{Have}
\frac{d\hat{H}}{dt}=\frac{\dot{\omega}}{\omega}(\hat{H}-\hat{L})+\gamma_a\omega^2(\hat{H}-\hat{L}).
\end{equation}
This is an equation for quantum mechanical operators and if we take the expectations we obtain Eq. (\ref{E}). Working analogously, we can find the Stratonovich differentials for $\hat{L}, \hat{C}$
\begin{eqnarray}
\label{Lang}\dbar \hat{L}&=&\left[-2\omega\hat{C}-\frac{\dot{\omega}}{\omega}(\hat{H}-\hat{L})\right]dt-\frac{\omega^2}{2}(\hat{p}\hat{q}+\hat{q}\hat{p})\dbar w_a\nonumber\\
             &&-\omega^2(\hat{p}\hat{q}+\hat{q}\hat{p})\dbar w_p,\\
\label{Cor}\dbar \hat{C}&=&\left(2\omega\hat{L}+\frac{\dot{\omega}}{\omega}\hat{C}\right)dt-m\omega^3\hat{q}^2\dbar w_a\nonumber\\
             &&+\omega\left(\frac{\hat{p}^2}{m}-m\omega^2\hat{q}^2\right)\dbar w_p,
\end{eqnarray}
and the corresponding It\={o} differentials
\begin{eqnarray}
\label{LIto}d\hat{L}&=&\left[-2\omega\hat{C}-\frac{\dot{\omega}}{\omega}(\hat{H}-\hat{L})+\gamma_a\omega^2(\hat{H}-\hat{L})\right]dt\nonumber\\
                    &&-4\gamma_p\omega^2\hat{L}dt-\omega\hat{C}dw_a-2\omega\hat{C}dw_p,\\
\label{CIto}d\hat{C}&=&\left(2\omega\hat{L}+\frac{\dot{\omega}}{\omega}\hat{C}-4\gamma_p\omega^2\hat{C}\right)dt\nonumber\\
                    &&-\omega(\hat{H}-\hat{L})dw_a+2\omega\hat{L}dw_p,
\end{eqnarray}
where note that $dw_p\cdot dw_p=2\gamma_pdt$. If we take in the last two equations the average over the noise, and in the resultant deterministic equations the expectation values of the operators, we recover Eqs. (\ref{L}) and (\ref{C}).

\section{EFFICIENCY BOUND IN THE PRESENCE OF PURE DEPHASING}

\label{app:bound}

We present first the case without noise, which will guide us in the case where only dephasing is present. In the absence of any noise mechanism, equations (\ref{x1}), (\ref{x2}) and (\ref{x3}) become
\begin{eqnarray}
\label{x1_a}\dot{x}_1&=&2x_3,\\
\label{x2_a}\dot{x}_2&=&-2ux_3,\\
\label{x3_a}\dot{x}_3&=&-ux_1+x_2.
\end{eqnarray}
Under the above evolution and for the initial conditions (\ref{init_cond}) it is $x_1x_2-x_3^2=1$ (Casimir invariant). At the final point, where (\ref{final_cond}) also holds, we find
\begin{equation}
\label{final_point}
x_1(T)=\frac{\omega_h}{\omega_c},\quad x_2(T)=\frac{\omega_c}{\omega_h}.
\end{equation}
This is the minimum achievable value of $x_2$, corresponding to minimum $E_c$ (\ref{min_energy}) and $\delta=0$. One way to obtain this value is in the adiabatic (long time) limit. Since the variables $x_1,x_2$ interact through $x_3$ in (\ref{x1_a}), (\ref{x2_a}), we first build a small positive value $x_3=\epsilon>0$ by applying $u=\omega_c^2/\omega_h^2$ for a sufficiently small time interval $dt\approx \epsilon/(1-\omega_c^2/\omega_h^2)$. It is not hard to see that $x_1(dt)>1$ and $x_2(dt)<1$. Subsequently, we maintain $x_3=\epsilon$ by applying the feedback control $u=x_2/x_1$, which gives $\dot{x}_3=0$ in (\ref{x3_a}). Note that $u(dt)<1$, thus (\ref{u_conditions}) is initially satisfied. During the application of this feedback control, the state equations become
\begin{eqnarray}
\label{x1_f}\dot{x}_1&=&2\epsilon,\\
\label{x2_f}\dot{x}_2&=&-2u\epsilon,\\
\label{x3_f}\dot{x}_3&=&0.
\end{eqnarray}
Observe that $x_1$ increases while $x_2$ decreases, thus $u$ also decreases from the value $u(dt)$ which is slightly less than unity. We apply the feedback law until $u$ reaches the lowest allowed bound $u=x_2/x_1=\omega_c^2/\omega_h^2$. Note that throughout this evolution it is $x_1x_2=1+\epsilon^2$. So, at the final point we find $x_1(T)=\omega_h\sqrt{1+\epsilon^2}/\omega_c, x_2(T)=\omega_c\sqrt{1+\epsilon^2}/\omega_h$. In the adiabatic limit $\epsilon\rightarrow 0$, it is $x_1(T)\rightarrow\omega_h/\omega_c, x_2(T)\rightarrow\omega_c/\omega_h$ and $x_3(T)\rightarrow 0$.

We now move to the pure dephasing case, where $\gamma_a=0$ and $\gamma_p>0$. Equations (\ref{x1}), (\ref{x2}) and (\ref{x3}) take the form
\begin{eqnarray}
\label{x1_de}\dot{x}_1&=&-2\gamma_pux_1+2\gamma_px_2+2x_3,\\
\label{x2_de}\dot{x}_2&=&2\gamma_pu^2x_1-2\gamma_pux_2-2ux_3,\\
\label{x3_de}\dot{x}_3&=&-ux_1+x_2-4\gamma_pux_3.
\end{eqnarray}
Again, we apply $u=\omega_c^2/\omega_h^2$ for $dt\approx \epsilon/(1-\omega_c^2/\omega_h^2)$ to create $x_3=\epsilon>0$. Then, we apply the feedback law $u=x_2/(x_1+4\gamma_px_3)$ and the above equations become
\begin{eqnarray}
\label{x1_feed}\dot{x}_1&=&(8\gamma_p^2u+2)\epsilon,\\
\label{x2_feed}\dot{x}_2&=&-u(8\gamma_p^2u+2)\epsilon,\\
\label{x3_feed}\dot{x}_3&=&0.
\end{eqnarray}
Observe again that $x_1$ increases while $x_2$ decreases, thus $u$ also decreases, and we apply this control until $u=x_2/(x_1+4\gamma_px_3)=\omega_c^2/\omega_h^2$.
Note that during the application of the feedback law it is $x_1x_2+4\gamma_p\epsilon x_2=c(\epsilon)$, a constant with limiting value $c(\epsilon)\rightarrow 1$ for $\epsilon\rightarrow 0$. At the final point we find $x_1(T)=\omega_h\sqrt{c(\epsilon)}/\omega_c-4\gamma_p\epsilon, x_2(T)=\omega_c\sqrt{c(\epsilon)}/\omega_h$. In the adiabatic limit $\epsilon\rightarrow 0$, it is $x_1(T)\rightarrow\omega_h/\omega_c, x_2(T)\rightarrow\omega_c/\omega_h$ and $x_3(T)\rightarrow 0$, thus the maximum efficiency is obtained. Note that this control strategy, to make a transfer between two variables through an intermediate variable which is kept small, has been used for the spin-order transfer along an Ising spin chain \cite{Stefanatos05}. The situation is reminiscent of STIRAP in a $\Lambda$-type atom, where perfect population transfer is achieved between two ground states coupled through a lossy excited state, which is actually never populated in the adiabatic (long time) limit \cite{Yuan12,Assemat12}. From (\ref{old_var}) we find that during the application of the feedback law it is $C/E_h=\sqrt{u}\epsilon<\epsilon$, while for $\epsilon\rightarrow 0$ it is additionally $L/E_h\approx 2\gamma_p\epsilon x_2/x_1<2\gamma_p\epsilon$. Thus, both $L$ and $C$ are of the order of $\epsilon$ or less, and the trajectory stays close to the noise-free subspace $(E,0,0)$.

\end{document}